\documentclass{pprai}

\usepackage{amsmath}
\usepackage[utf8]{inputenc}
\usepackage[T1]{fontenc}
\usepackage{graphicx}
\usepackage{txfonts}
\usepackage{url}
\usepackage{algpseudocode}

\usepackage[final]{changes}

\title{Task scheduling for autonomous vehicles \\ in the Martian environment}
\headtitle{Task scheduling for autonomous vehicles in the Martian environment}

\author{Wojciech Burzyński$^{1[0009-0003-8668-7754]}$, Mariusz Kaleta$^{2[0000-0002-2225-8956]}$}
\headauthor{W. Burzyński, M. Kaleta}
\affiliation{%
  $^1$Warsaw University of Technology\\
  Doctoral School\\
  Pl. Politechniki 1, 00-661 Warsaw, Poland\\
  wojciech.burzynski.dokt@pw.edu.pl\\
  $^2$Warsaw University of Technology\\
  Faculty of Electronics and Information Technology\\
  Nowowiejska 15/19, 00-665 Warsaw, Poland\\
  mariusz.kaleta@pw.edu.pl\\}

\keywords{task scheduling, electric VRP/TSP, autonomous vehicles, mars exploration}

\begin{document}
\maketitle

\begin{abstract}
In the paper, we introduced a novel variant of Electric VRP/TSP, the Solar Powered Rover Routing Problem (SPRRP), to tackle the routing of energy-constrained autonomous electric vehicles for Martian missions. We proposed a basic formulation of the problem based on the graph model that decomposes each Point of Interest into movement, charging, and research tasks. We have also outlined further possibilities for extending the problem.
\end{abstract}

\section{Introduction} \label{introduction}

A Mars exploration mission is a complex undertaking that requires detailed planning, integration of the newest technologies, and proper budget management. In general, related decisions that have to be made can be divided into strategic, tactical, and operational kinds. The strategic level concerns aim, definition, budget preparation, fleet composition, and strategic resource planning. The tactical level encompasses all decisions related to choosing and planning research tasks to be done within a certain horizon (e.g., weeks or months). At the operational level, tactical plans are put into action. However, due to high uncertainty and burdens in communication, exploration of extraterrestrial planets requires the use of highly autonomous vehicles. Even though the only mission so far that used multiple vehicles cooperating was \emph{Mars 2020} with rover \emph{Perseverance} and helicopter \emph{Ingenuity} serving as a scout, authors believe that future missions will require cooperation and scheduling of multiple autonomous vehicles. In this paper, we focus solely on the tactical level, assuming that strategic decisions constitute an input to tactical decision models, while the results of tactical models become inputs or may be used to update the aims and tasks of autonomous vehicles at the operational level.

Most of the literature discussing Mars exploration focuses on the operational level, including modeling of kino-dynamic properties of rovers, mechanical properties of Martian soil, computer vision, and reactive trajectory planning. In \cite{Hedrick2020TerrainAwarePP}, authors propose a way to evaluate terrain based on satellite imaging and integrate it into a trajectory planner. The problem of maximizing scientific data acquisition by robots in space exploration missions is considered in \cite{Colby}. The authors proposed a human-in-the-loop multi-agent learning system in which a human team on the Earth identifies a set of Points of Interest (PoIs) and delegates tasks to autonomous agents. In \cite{Huang}, tasks are assigned to robots in a swarm by scientists, and each robot applies its policy to perform its tasks on the simulated Mars surface. The control architecture for multiple-robot planetary outposts (CAMPOUT) has been proposed in \cite{Huntsberger}. This framework is devoted to cooperating with multiple robots performing tightly coordinated tasks in unpredictable terrain.

In the literature, the scheduling and routing of vehicles were described in terms of the Travelling Salesman Problem (TSP) and the Vehicle Routing Problem (VRP). Their objective is to visit a given set of customers with a vehicle or a fleet of vehicles while satisfying various constraints. In the case of the Martian environment, vehicles visit interesting points to perform research tasks such as soil analysis using a spectrometer. For a comprehensive overview of VRP literature, the reader is referred to \cite{ELSHAER2020106242}, and for VRP variants concerning electric vehicles, to \cite{kucukoglu}.

This paper proposes a novel variant of Electric VRP/TSP, that is, the Solar Powered Rover Routing Problem (\textbf{SPRRP}), to tackle the routing of energy-constrained electric vehicles that are powered by solar panels such as Martian rovers. Contrary to existing variants, vehicles do not have access to charging stations, and their energy consumption depends on what the vehicle is doing and the time of \emph{sol} (martian day). According to the author's best knowledge, such a variant has not yet been discussed in the literature.

\section{Problem and model definition} \label{problem}

\subsection{Solar Powered Rover Routing Problem (SPRRP)}

\begin{figure}[th]
\centering
\includegraphics[width=\textwidth]{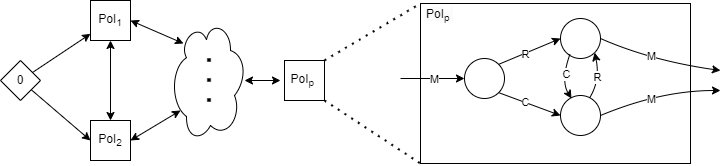}
\caption{The graph model of the problem. The left part is a flow network representing PoIs with starting node 0, and the right part shows a decomposition of PoI $p$ into specific tasks.}
\label{fig:sprrp}
\end{figure}

The task is to plan routes for autonomous vehicles visiting Points of Interest (PoIs); however, some PoIs may remain not visited. Let $Z=\{M, R, C\}$ be the set of task types, where $M, R, C$ means \emph{Movement}, carrying out the \emph{Research}, and \emph{Charging}, respectively. A vehicle travels between PoIs (\emph{M}), performs the research (\emph{R}), and optionally can charge (\emph{C}) at each PoI. We developed a graph model presented in Fig. \ref{fig:sprrp}, in which each PoI is modeled with three nodes that allow research and charging to be done in any order at PoI. We assume that all PoIs are connected. In other words, for each PoI, there are edges representing \emph{Movement} task leading to an entry node of every other PoI. All vehicles start at node 0. In the graph model $G=(V, E)$, a task is represented by edge $(i,j)\in E$ and is of type $z_{ij} \in Z$, and node $v\in V$ represents an event related to starting or ending tasks. Each task must be assigned to the vehicle $k \in K$ and is described by starting time $t^k_{ij}$ and duration $\tau^k_{ij}$ (we assume constant charging time).

The SPRRP problem can be formulated as a non-linear programming model as follows:
\begin{equation}
    \max \sum_{(i,j) \in E}{\sum_{k \in K}{c_{ij}x_{ij}^k}} \label{sprrp:obj}
\end{equation}
subject to
\begin{align}
& \sum_{i \in V}x^k_{ij}=y^k_j && \forall j \in V\backslash \{ 0 \}, \forall k \in K \label{one_enters}\\
& \sum_{i \in V}x^k_{ji}=y^k_j && \forall j \in V, \forall k \in K \label{one_leaves}\\
& \sum_{k \in K}{y^k_i} \leq 1 && \forall i \in V\backslash \{ 0 \} \label{each_at_most}\\
& b^k_j \leq \sum_{i \in V}{x^k_{ij}(b^k_i + \Delta E(t^k_i, \tau^k_{ij}, z_{ij}))} &&  \forall j \in V \backslash \{0\}, \forall k \in K \label{battery}\\
& 0 \leq b^k_i \leq B && \forall i \in V, \forall k \in K \label{battery_limits}\\
& t^k_0 = T_0 && \forall k \in K \label{initial_time}\\
& t^k_j = \sum_{i \in V}{x^k_{ij}(t^k_i +\tau_{ij})} && \forall j \in V \backslash \{0\}, \forall k \in K \label{time} \\
& 0 \leq t^k_j \leq T_{max} && \forall k \in K, \forall j \in V \label{tmax} \\
& y_j^k \in \{0,1\}, x^k_{ij} \in \{0,1\} && \forall k \in K, \forall i,j \in V  \label{sprrp:vars}
\end{align}
where $x^k_{ij}$ is a binary variable that equals to 1 if vehicle $k$ traverses edge $(i,j)$, binary variable $y_j^k$ is 1 if vehicle $k$ visits PoI $j$,
$b^k_i$ is battery energy of vehicle $k$ at event $i$, and $t^k_i$ is the time of occurrence of the event $i$ and vehicle $k$.

The objective (\ref{sprrp:obj}) maximizes benefits from completing tasks, where $c_{ij}$ is the benefit assigned to task $(i,j)$. Constraints (\ref{one_enters})-(\ref{one_leaves}) ensure that if a vehicle is assigned to event $j$ it must enter and leave node $j$. Constraint (\ref{each_at_most}) ensures that each node is visited at most once. Note that this assumption does not exclude the case in which a vehicle can physically pass through the PoI previously visited; however, in the graph, there will be a direct connection between the origin and destination. Tracking the battery level $b_j^k$ is done by constraints (\ref{battery})-(\ref{battery_limits}), where $B$ is battery capacity and $b_0^k$ is initial battery energy. Constraints (\ref{initial_time})-(\ref{tmax}) track the time of event $j$, where $T_0$ and $T_{max}$ are the initial moment and the time limit on the mission, respectively.

Due to constraints (\ref{one_enters}) and (\ref{one_leaves}) (path consistency)  and (\ref{time}) (subtour elimination), a vehicle must reach any PoI along some path to be able to complete research task at that PoI, and only completion of research tasks increases the objective. Moreover, constraints (\ref{battery}) and (\ref{battery_limits})  force a vehicle to traverse an edge equivalent to pure charging if, otherwise, it cannot perform research tasks due to lack of power.

\subsection{Energy model} \label{energy_model}

Let $P(z,t) = P_+(z,t) + P_-(z,t), z \in Z, t \in \mathbb{R}$ be the total power of the vehicle, where $z$ is a type of task performed by a vehicle at time $t$. $P_-(z,t)$ is the power consumed due to $z$-type task realization and $P_+(z,t)$ is power gain from solar panels. The energy cost of a task type $z$ starting at $t$ and lasting $\tau$ is defined as follows:
\begin{equation}
    \Delta E(t,\tau, z) = \int^{t+\tau}_{t}{P(z,t)\mathrm{d}t}
\end{equation}

To simplify the notation and without loss of the generality, we assume simplified power functions as follows:
\begin{align} \label{eq:p-}
P_-(z,t) &= \begin{cases}
                -a & z = M \\
                -b & z = R \\
                -c & z = C
            \end{cases},\\
P_+(z,t) &= \frac{sgn(sin(\frac{2 \pi t}{T}))}{2} + \frac{1}{2}, \quad T = 1 \text{[sol]}, \label{eq:p+}
\end{align}
where $a, b, c$ are constant power costs defined for each type of task. Introducing different power costs for each task only requires an additional index in the formula~(\ref{eq:p-}). $P_+(z,t)$ is defined as a square wave with a period of one sol.
This simple model neglects several aspects, including dependence on time of day, weather, heating-up phenomena, etc. Ignoring the position of a given rover can be justified if distances between PoIs are not high, so the differences in the weather and other factors can be neglected.

\section{Numerical example}

\begin{figure}[th]
\centering
\includegraphics[width=\textwidth]{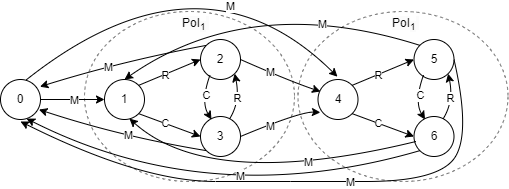}
\caption{The graph model of the test problem.}
\label{fig:graph_example}
\end{figure}

\begin{figure}[th]
\centering
\includegraphics[width=0.5\textwidth]{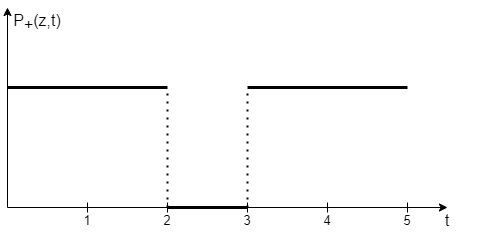}
\caption{Function of power $P_{+}(z,t)$ in the numerical example.}
\label{fig:power_function}
\end{figure}

To check the validity of the model, we prepared a simple test case of one vehicle and two PoIs. The graph model of the problem is presented in Fig. \ref{fig:graph_example}. Each task can be performed in one unit of time ($\tau_{ij}=1$), and Table \ref{tab:energy_cost} presents the energy consumption for each task. The value of each task is $c_{ij}=1$. The definition of $P_{+}(z,t)$ is presented in Fig \ref{fig:power_function}.

\begin{table}[!ht]
\caption{Energy consumption}\label{tab:energy_cost}
\centering
\begin{tabular}{|l|r|r|}
\hline
\textbf{task type} & \textbf{PoI$_1$} & \textbf{PoI$_2$} \\
\hline\hline
M Base-PoI & 6 & 10 \\
M PoI-PoI & 4 & 4 \\
R & 5 & 5 \\
C & 1 & 1\\
\hline
\end{tabular}
\end{table}

To solve the model (\ref{sprrp:obj})-(\ref{sprrp:vars}) we linearized constraints (\ref{battery}) and (\ref{time}) by introducing additional variables, including binary variables, to represent $\Delta E(t^k_i, \tau^k_{ij}, z_{ij}))$ as a piecewise function and to model product of variables. We used CPLEX to solve the resulting Mixed Integer Problem (MIP).

To observe the behavior of the optimal results, we manipulated the battery capacity installed on the vehicle. For a battery capacity equal to 4, the optimal solution is to visit only PoI$_1$ and to plan the charging task before performing the research at PoI$_1$. However, when the capacity is increased to 5, the optimal solution still includes only PoI$_1$, but no charging task is planned in that case. With battery capacity ranging from 6 to 10, the model results in visiting both PoIs, but charging task at PoI$_1$ is required. Finally, with a capacity greater or equal to 11, the vehicle is scheduled to serve both PoIs without additional charging tasks. The model behaves correctly and we showed that it can be solved by MIP solvers after linearization. However, since it requires additional variables and constraints, it may suffer from the complexity in case of bigger problems.

\section{Problem variants and extensions}

Various variants and extensions of SPRRP can be formulated.
To grasp the multidimensionality of the problem,
we propose the following dimensions describing it:

\begin{itemize}
  \item vehicles, e.g., capacitated, heterogeneous fleet (specialized vehicles), different energy cost functions;
  \item tasks/PoIs, e.g., with time windows (for instance, pictures must be taken in specific light conditions), deterministic/stochastic time (task execution and energy budget can be uncertain), obligatory visit of each PoI, complex tasks at PoIs;
  \item environment, e.g., nondeterministic or dynamic traveling times (best satellite imaging on Mars yields a resolution of 1m/pixel \cite{Hirise} and covers only a small fraction of Mars's surface);
  \item objective, e.g., distance, time, cost, delays, benefits, risks, including multi-criteria versions;
\end{itemize}

\section{Summary} \label{conclusions}

We introduced and formulated a new problem, the Solar Powered Rover Routing Problem, similar to the well-known class of VRP problems but better suited to future Martian missions. Future research should be focused on providing algorithms and numerical results for the introduced model. We also plan to test the model combined with operational algorithms with real vehicles in a simulated environment on Earth. We also presented possible variants of the problem, demonstrating the plethora of the issue for future research work.

\end{document}